\documentclass[12pt]{article}
\usepackage{amsfonts}
\usepackage{amssymb}
\usepackage{amsmath}
\newcommand{\eqdef}{\stackrel{\text{def}}{=}}

\newcommand{\ignore}[1]{}

\begin{document}

\baselineskip=20pt
\renewcommand{\theequation}{\arabic{section}.\arabic{equation}}

%%%%%%%%%%%%%%%%%%%%%%%%%%%%%%%%%%%%%%%%%%%%%%%%%%%%%%%%%%%%
%                                                          %
%  Title page                                              %
%                                                          %
%%%%%%%%%%%%%%%%%%%%%%%%%%%%%%%%%%%%%%%%%%%%%%%%%%%%%%%%%%%%
\newfont{\elevenmib}{cmmib10 scaled\magstep1}
\newcommand{\preprint}{
     %\vspace*{-20mm}\begin{flushleft} Version: 6 August 2007 \end{flushleft}
     \begin{flushleft}
       \elevenmib Yukawa\, Institute\, Kyoto\\
     \end{flushleft}\vspace{-1.3cm}
     \begin{flushright}\normalsize \sf
       DPSU-07-3\\
       YITP-07-44\\
       August 2007
     \end{flushright}}
\newcommand{\Title}[1]{{\baselineskip=26pt
     \begin{center} \Large \bf #1 \\ \ \\ \end{center}}}
\newcommand{\Author}{\begin{center}
     \large \bf Satoru Odake${}^a$ and Ryu Sasaki${}^b$ \end{center}}
\newcommand{\Address}{\begin{center}
       $^a$ Department of Physics, Shinshu University,\\
       Matsumoto 390-8621, Japan\\
       ${}^b$ Yukawa Institute for Theoretical Physics,\\
       Kyoto University, Kyoto 606-8502, Japan
     \end{center}}
\newcommand{\Accepted}[1]{\begin{center}
     {\large \sf #1}\\ \vspace{1mm}{\small \sf Accepted for Publication}
     \end{center}}

\preprint
\thispagestyle{empty}
\bigskip\bigskip\bigskip

\Title{Multi-Particle Quasi Exactly Solvable Difference Equations}
\Author

\Address
\vspace{1cm}

\begin{abstract}
Several explicit examples of
{\em multi-particle quasi exactly solvable\/} `discrete' quantum
mechanical Hamiltonians are derived by deforming the well-known
exactly solvable multi-particle Hamiltonians,
the Ruijsenaars-Schneider-van Diejen systems.
These are difference analogues of the quasi exactly solvable
multi-particle systems, the quantum Inozemtsev systems obtained by
deforming the well-known exactly solvable Calogero-Sutherland systems.
They have a finite number of exactly calculable eigenvalues and
eigenfunctions. This paper is a multi-particle extension of
the recent paper by one of the authors on deriving quasi exactly
solvable {\em difference\/} equations of single degree of freedom.
\end{abstract}

%%%%%%%%%%%%%%%%%%%%%%%%%%%%%%%%%%%%%%%%%%%%%%%%%%%%%%%%%%%%%%%
%                                                             %
%  1. Introduction                                            %
%                                                             %
%%%%%%%%%%%%%%%%%%%%%%%%%%%%%%%%%%%%%%%%%%%%%%%%%%%%%%%%%%%%%%%
\section{Introduction}
\label{intro}
\setcounter{equation}{0}

Recently a recipe to obtain a quasi exactly solvable {\em difference\/}
equation from an exactly solvable difference equation is developed
by one of the present authors \cite{deltaqes}.
In the present paper we apply the recipe to obtain multi-particle
quasi exactly solvable difference equations.
A quantum mechanical system is called Quasi Exactly Solvable (QES),
if a finite number of eigenvalues and the corresponding eigenfunctions
can be determined exactly \cite{Ush}.
Since the number of exactly solvable states can be chosen as
large as wanted, a QES system could be used as a good alternative to
an exactly solvable system for theoretical as well as practical purposes.
Many examples of QES systems of single degree of freedom have been
known for some time, whereas the known examples of those with many
degrees of freedom are rather limited in their structure \cite{st1,turb2}.
They are all obtained by deforming Calogero-Sutherland systems
\cite{cal,sut,kps}, the exactly solvable Hamiltonian dynamics based
on root systems. Among them those based on the $BC$ type root systems
(rational and trigonometric) and on the $A$ type root systems
(trigonometric) can contain an arbitrary number of particles.
Therefore the examples of multi-particle QES are infinite in number.
In the present paper we derive several examples of multi-particle QES
{\em difference\/} equations by deforming Ruijsenaars-Schneider van Diejen
(RSvD) systems \cite{RS,vD,os2,os6}, which are the difference equation
analogues of Calogero-Sutherland systems.
To be more precise, we derive multi-particle QES difference equations
by deforming RSvD systems based on $BC$ (rational and trigonometric) and
$A$ (trigonometric) type root systems.
Thus the examples of multi-particle QES difference equations are now
infinite in number. The Hamiltonian of RSvD systems has two kinds of
interaction terms, the `single particle interaction' part and the
`multi-particle interaction' part. The latter is kept intact and
the former, the `single particle interaction' part is {\em deformed\/}
according to the recipe given in the recent paper \cite{deltaqes}.

This paper is organised as follows.
In section \ref{rat} the Hamiltonian of the rational $BC$ type theory
is derived and the finite dimensional invariant polynomial space is
identified. The trigonometric $BC$ type theory is explained in section
\ref{trig}. The Hamiltonian of the trigonometric $A$ type theory and
the finite dimensional invariant polynomial space are derived in section
\ref{trigA}. The final section is for a short summary and comments.

%%%%%%%%%%%%%%%%%%%%%%%%%%%%%%%%%%%%%%%%%%%%%%%%%%%%%%%%%%%%%%%
%                                                             %
%  2. Rational $BC$ type theory                               %
%                                                             %
%%%%%%%%%%%%%%%%%%%%%%%%%%%%%%%%%%%%%%%%%%%%%%%%%%%%%%%%%%%%%%%
\section{Rational $BC$ type theory}
\label{rat}
\setcounter{equation}{0}

The first multi-particle QES Hamiltonian is a simple deformation of
the rational Ruijsenaars-Schneider-van Diejen (RSvD) \cite{RS,vD}
system based on the $BC_n$ root system. Here $n$ is the rank of
the root system as well as the degree of freedom with coordinates
and conjugate momenta:
\[
  x\eqdef(x_1,x_2,\ldots,x_n)\in\mathbb{R}^n,\quad
  p\eqdef(p_1,p_2,\ldots,p_n).
\]
As always, the momentum operator is realised as a differential operator
\(
  p_j=-i\partial_j=-i\partial/\partial x_j.
\)
In `discrete' quantum mechanical Hamiltonians the momentum operators
appear in exponentiated forms \(e^{\pm p_j}=e^{\mp i\partial_j}\),
in contrast to the ordinary quantum mechanics, in which the momentum
operators appear as polynomials.
Thus their action on the wavefunction is a finite shift in the imaginary
direction:
\[
  e^{\pm p_j}\psi(x)=\psi(x_1,\ldots,x_{j-1},x_j\mp i,x_{j+1},\ldots,x_n),
\]
leading to {\em difference} Schr\"odinger equations.
The quasi exactly solvable rational Hamiltonian has the following
general form \cite{deltaqes,os4os5,os6}:
\begin{align}
  \mathcal{H}\,&\eqdef
  \sum_{j=1}^n\Bigl(\sqrt{V_j}\,e^{\, p_j}\sqrt{V_j^*}
  +\sqrt{V_j^*}\,e^{-p_j}\sqrt{V_j}-V_j-V_j^*\Bigr)
  +\alpha_{\mathcal{M}},\quad \mathcal{M}\in\mathbb{N},\\
  &=\sum_{j=1}^n\Bigl(\sqrt{V_j}\,e^{-i\partial_j}\sqrt{V_j^*}
  +\sqrt{V_j^*}\,e^{\,i\partial_j}\sqrt{V_j}-V_j-V_j^*\Bigr)
  +\alpha_{\mathcal{M}}.
\end{align}
Here \(\alpha_{\mathcal{M}}\) is a compensation term indexed by
a natural number $\mathcal{M}$, to be specified shortly in
\eqref{type1w} and \eqref{type2al}.
The potential function $V_j$ has the following general form:
\begin{equation}
  V_j(x)\eqdef w(x_j)\prod_{\genfrac{}{}{0pt}{}{k=1}{k\neq j}}^n
  \prod_{\varepsilon=\pm 1}v(x_j+\varepsilon x_k),
  \label{ratvj}
\end{equation}
in which the `multi-particle interaction' part $v$ is
{\em not deformed\/}
\begin{equation}
  v(y)\eqdef 1-i\,\frac{g}{y},\quad g>0.
\end{equation}
Whereas the `single particle interaction' part $w$ allows
{\em two types of deformation\/} for QES, corresponding to
the linear and quadratic polynomial deformations introduced
in section 3.1 of \cite{deltaqes}:
\begin{alignat}{3}
  &\text{Type I}&&:&\quad w(y)&\eqdef(a_2+i y)w_0(y),\quad
  \alpha_\mathcal{M}(x)\eqdef\mathcal{M}\sum_{j=1}^nx_j^2,
  \label{type1w}\\
  &\text{Type II}&&:& w(y)&\eqdef(a_1+i y)(a_2+i y)w_0(y),
  \label{type2w}\\
  &&&& \alpha_\mathcal{M}(x)&\eqdef
  \mathcal{M}\Bigl(\mathcal{M}-1+\sum_{\alpha=1}^6a_\alpha+2(n-1)g\Bigr)
  \sum_{j=1}^nx_j^2,
  \label{type2al}
\end{alignat}
with a common undeformed $w_0(y)$ \cite{os4os5,os6}
\begin{equation}
  w_0(y)\eqdef\frac{\prod_{\alpha=3}^6(a_\alpha+i y)}{2iy(2iy +1)},
  \quad a_\alpha>0.
  \label{w0Wilson}
\end{equation}
As a single particle dynamics, the above $w_0$ corresponds to the
deformation of the harmonic oscillator with the centrifugal potential.
The corresponding eigenfunctions are the Wilson polynomials
\cite{os4os5,askey}.

The main part of the Hamiltonian is factorised \cite{deltaqes,os4os5,os6}:
\begin{align}
  \mathcal{H}&=
  \sum_{j=1}^n\mathcal{A}_j^{\dagger}\mathcal{A}_j+\alpha_{\mathcal{M}},\\
  \mathcal{A}_j&\eqdef -i\left(e^{-\frac{i}{2}\partial_j}\sqrt{V_j^*}
  -e^{\frac{i}{2}\partial_j}\sqrt{V_j}\right),\quad
  \mathcal{A}_j^{\dagger}=i\left(\sqrt{V_j}\,e^{-\frac{i}{2}\partial_j}
  -\sqrt{V_j^*}\,e^{\frac{i}{2}\partial_j}\right),%\quad j=1,\ldots,n.
\end{align}
exhibiting the hermiticity (self-adjointness) of the Hamiltonian.

The {\em pseudo ground state\/} wavefunction $\phi_0$ is defined as
the one annihilated by all the $\mathcal{A}_j$ operators:
\begin{equation}
  \mathcal{A}_j\phi_0=0,\quad j=1,\ldots, n.
\end{equation}
It is given by
\begin{alignat}{3}
  &\text{Type I}&&:&\quad \phi_0(x)&\eqdef\biggl|\,\prod_{j=1}^n
  \frac{\prod_{\alpha=2}^6\Gamma(a_{\alpha}+ix_j)}{\Gamma(2ix_j)}\,\cdot\!\!
  \prod_{1\leq j<k\leq n}\prod_{\varepsilon=\pm1}
  \frac{\Gamma(g+i(x_j+\varepsilon x_k))}{\Gamma(i(x_j+\varepsilon x_k))}
  \,\biggr|,\\
  &\text{Type II}&&:& \phi_0(x)&\eqdef\biggl|\,\prod_{j=1}^n
  \frac{\prod_{\alpha=1}^6\Gamma(a_{\alpha}+ix_j)}{\Gamma(2ix_j)}\,\cdot\!\!
  \prod_{1\leq j<k\leq n}\prod_{\varepsilon=\pm1}
  \frac{\Gamma(g+i(x_j+\varepsilon x_k))}{\Gamma(i(x_j+\varepsilon x_k))}
  \,\biggr|,
\end{alignat}
in which an abbreviation $|f|\eqdef\sqrt{ff^*}$ is used.
It is obvious that these $\phi_0$ are square integrable in the
principal Weyl chamber of $BC_n$:
\begin{equation}
  \int_{PW}\phi_0(x)^2d^{\,n}x<\infty,\qquad
  PW\eqdef\bigl\{x\in\mathbb{R}^n\bigm|x_1>x_2>\cdots>x_n>0\bigr\},
  \label{priwc}
\end{equation}
and they have no node or singularity in $PW$.
By the similarity transformation in terms of the above pseudo ground
state wavefunction $\phi_0(x)$, we obtain a Hamiltonian
$\tilde{\mathcal H}$ leading to a finite difference eigenvalue equation
with rational potentials \eqref{ratvj}--\eqref{w0Wilson}
\cite{deltaqes,os4os5}:
\begin{align}
  &\tilde{\mathcal{H}}\eqdef\phi_0^{-1}\circ\mathcal{H}\circ\phi_0
  =\sum_{j=1}^n
  \Bigl(V_j(x)(e^{-i\partial_j}-1)+V_j(x)^*(e^{\,i\partial_j}-1)\Bigr)
  +\alpha_{\mathcal{M}}(x),\\
  &\mathcal{H}\phi=\mathcal{E}\phi,\quad
  \phi(x)=\phi_0(x)P_{\mathcal{M}}(x)\quad
  \Longleftrightarrow\quad
  \tilde{\mathcal{H}}P_{\mathcal{M}}=\mathcal{E}P_{\mathcal{M}}.
  \label{eigfun}
\end{align}

In the {\em undeformed\/} limit, {\em i.e.} $w=w_0$ and
$\alpha_{\mathcal M}=0$, the theory is the exactly solvable rational
$BC_n$ RSvD \cite{vD} system with a Hamiltonian $\tilde{\mathcal H}_0$.
The corresponding $\phi_0$ becomes the true ground state wavefunction.
The exact solvability means that $\tilde{\mathcal H}_0$ maps a
Weyl-invariant polynomial in $\{x_j\}$ into another of the same degree.
The $BC_n$ Weyl-invariant polynomials in $\{x_j\}$ are simply symmetric
(under any permutation $j\leftrightarrow k$) polynomials in $\{x_j^2\}$.
For later convenience, let us introduce a monomial symmetric polynomial
\begin{equation}
  m_{\lambda}(\{y_j\})
  =m_{(\lambda_1,\ldots,\lambda_n)}(y_1,\ldots,y_n)
  \eqdef\sum_{(l_1,\ldots,l_n)}y_1^{l_1}\cdots y_n^{l_n},
\end{equation}
where the summation with respect to $(l_1,\ldots,l_n)$ is taken over
all distinct permutations of $\lambda\eqdef(\lambda_1,\dots,\lambda_n)$.

In the {\em deformed\/} theory, it is straightforward to demonstrate
that $\tilde{\mathcal H}$ maps a symmetric polynomial in $\{x_j^2\}$
of degree equal or less than $\mathcal{M}$ into another:
\begin{align}
  \tilde{\mathcal{H}}\,\mathcal{V}_{\mathcal{M}}&\subseteq
  \mathcal{V}_{\mathcal{M}},
  \qquad\qquad\qquad\qquad
  \dim\mathcal{V}_{\mathcal{M}}=\genfrac{(}{)}{0pt}{1}{\mathcal{M}+n}{n},
  \label{subspac1}\\[4pt]
  \mathcal{V}_{\mathcal{M}}&\eqdef\text{Span}\bigl[
  m_{(l_1,\ldots,l_n)}(\{x_j^2\})\bigm|
  0\leq l_j\leq\mathcal{M},\ 1\leq j\leq n\bigr].
  \label{eq:V0def}
\end{align}
This establishes the quasi exact solvability.
The proof that an invariant polynomial is mapped into another goes
almost parallel with the undeformed theory.
One simply has to verify the vanishing of the residues of the simple
poles at $q_j=\pm q_k,0,\pm i/2$.
The other step is to show that a symmetric polynomial in $\{x_j^2\}$
with degree $m$ ($\leq\mathcal{M}-1$) is mapped to another with degree
$m+1$, whereas a symmetric polynomial of degree $\mathcal{M}$ remains
with the same degree.
This part goes almost the same as in the single particle case shown
in \cite{deltaqes}, since it is caused by the deformation of $w$,
the single particle interaction part.
The present examples are the difference equation version of the QES
theory called rational $BC$ Inozemtsev systems discussed in section
6 of Sasaki-Takasaki paper \cite{st1}.

%%%%%%%%%%%%%%%%%%%%%%%%%%%%%%%%%%%%%%%%%%%%%%%%%%%%%%%%%%%%%%%
%                                                             %
%  3. Trigonometric $BC$ type theory                          %
%                                                             %
%%%%%%%%%%%%%%%%%%%%%%%%%%%%%%%%%%%%%%%%%%%%%%%%%%%%%%%%%%%%%%%
\section{Trigonometric $BC$ type theory}
\label{trig}
\setcounter{equation}{0}

The next example is a QES deformation of the trigonometric $BC_n$
RSvD system. Because of the periodicity of the trigonometric potential,
we introduce a slightly different notation for the dynamical variables:
\begin{equation}
  \theta\eqdef (\theta_1,\theta_2,\ldots,\theta_n)\in\mathbb{R}^n,\quad
  0<\theta_j<\pi,\quad x_j=\cos\theta_j,\quad z_j=e^{i\theta_j}.
\end{equation}
The dynamical variables are $\theta$.
We denote $D_j\eqdef z_j\frac{d}{dz_j}$. Then $q^{D_j}$ is a $q$-shift
operator,
\[
  q^{D_j}f(z)=f(z_1,\ldots,z_{j-1},qz_j,z_{j+1},\ldots,z_n),
\]
with $0<q<1$.
The quasi exactly solvable trigonometric Hamiltonian has the following
general form \cite{deltaqes,os4os5,os6}:
\begin{equation}
  \mathcal{H}\eqdef\sum_{j=1}^n\Bigl(\sqrt{V_j}\,q^{D_j}\sqrt{V_j^*}
  +\sqrt{V_j^*}\,q^{-D_j}\sqrt{V_j}-V_j-V_j^*\Bigr)
  +\alpha_{\mathcal{M}}.\label{trigH}
\end{equation}
The compensation term $\alpha_{\mathcal{M}}$ is given in
\eqref{altrigBC}. The potential function $V_j$ consists of the
`single particle interaction' part $w$ and the `multi-particle
interaction' part $v$:
\begin{align}
  V_j(z)&\eqdef w(z_j)\prod_{\genfrac{}{}{0pt}{}{k=1}{k\neq j}}^n
  \prod_{\varepsilon=\pm 1}v(z_jz_k^{\varepsilon}),
  \label{vjz}\\
  v(y)&\eqdef\frac{1-a_0y}{1-y},\quad w(y)\eqdef(1-a_1y)w_0(y),\quad
  w_0(y)\eqdef\frac{\prod_{\alpha=2}^5(1-a_{\alpha}y)}{(1-y^2)(1-qy^2)},
  \label{wz}\\
  \alpha_{\mathcal{M}}(z)&\eqdef(q^{\mathcal{M}}-1)q^{-1}a_0^{2(n-1)}
  a_1a_2a_3a_4a_5\sum_{j=1}^n\bigl(z_j+z_j^{-1}\bigr).
  \label{altrigBC}
\end{align}
As before, the `multi-particle interaction' part $v$ is not deformed
but the `single particle interaction' part $w$ is multiplicatively
deformed by a linear term from $w_0$.
As a single particle dynamics, the above $w_0$ corresponds to the
deformation of the P\"oschl-Teller potential. The corresponding
eigenfunctions are the Askey-Wilson polynomials \cite{os4os5,askey}.

The main part of the Hamiltonian is factorised \cite{deltaqes,os4os5,os6}:
\begin{align}
  \mathcal{H}&
  =\sum_{j=1}^n\mathcal{A}_j^{\dagger}\mathcal{A}_j+\alpha_{\mathcal{M}},
  \label{trigfact}\\
  \mathcal{A}_j&\eqdef i\Bigl(q^{\frac12D_j}\sqrt{V_j^*}
  -q^{-\frac12D_j}\sqrt{V_j}\Bigr),\quad
  \mathcal{A}_j^{\dagger}=-i\Bigl(\sqrt{V_j}\,q^{\frac12D_j}
  -\sqrt{V_j^*}\,q^{-\frac12D_j}\Bigr),
  \label{trigAAdag}
\end{align}
and the {\em pseudo ground state\/} wavefunction $\phi_0$
\begin{equation}
  \phi_0(z)\eqdef\biggl|\,\prod_{j=1}^n
  \frac{(z_j^2\,;q)_{\infty}}
       {\prod_{\alpha=1}^5(a_{\alpha}z_j\,;q)_{\infty}}\,\cdot\!\!
  \prod_{1\leq j<k\leq n}\prod_{\varepsilon=\pm 1}
  \frac{(z_jz_k^{\varepsilon}\,;q)_{\infty}}
       {(a_0z_jz_k^{\varepsilon}\,;q)_{\infty}}
  \,\biggr|,
\end{equation}
is obtained as the common zero-mode of all $\mathcal{A}_j$ operators
$\mathcal{A}_j\phi_0=0$, $j=1,\ldots,n$.
Here the standard notation
$(a\,;q)_{\infty}\eqdef\prod_{n=0}^{\infty}(1-aq^n)$ and
$|f|\eqdef\sqrt{ff^*}$ are used.
It is obvious that $\phi_0$ has no zero or singularity in the principal
Weyl alcove of $BC_n$
\begin{equation}
  PW_T\eqdef\bigl\{\theta\in\mathbb{R}^n\bigm|\pi>\theta_1+\theta_2>
  \theta_1>\theta_2>\cdots >\theta_n>0\bigr\},
  \label{priwa}
\end{equation}
so long as the parameters are restricted to
\begin{equation}
  -1<a_\alpha<1,\quad \alpha=0,1,\ldots,5.
\label{trigbcparas}
\end{equation}
Then the square integrability of $\phi_0$,
$\int_{PW_T}\phi_0^2d^n\theta<\infty$ is trivial.
By the similarity transformation in terms of the above pseudo ground
state wavefunction $\phi_0(z)$, we obtain a Hamiltonian
$\tilde{\mathcal H}$ leading to a finite difference eigenvalue equation
with rational potentials in $z$ and $z^{-1}$ \eqref{vjz}--\eqref{wz}
\cite{deltaqes,os4os5}:
\begin{align}
  &\tilde{\mathcal{H}}\eqdef\phi_0^{-1}\circ\mathcal{H}\circ\phi_0
  =\sum_{j=1}^n
  \Bigl(V_j(z)(q^{\,D_j}-1)+V_j(z)^*(q^{-D_j}-1)\Bigr)
  +\alpha_{\mathcal{M}}(z),
  \label{trigHtilde}\\
  &\mathcal{H}\phi=\mathcal{E}\phi,\quad
  \phi(z)=\phi_0(z)P_{\mathcal{M}}(z)\quad
  \Longleftrightarrow\quad
  \tilde{\mathcal{H}}P_{\mathcal{M}}=\mathcal{E}P_{\mathcal{M}}.
  \label{trigeigeneq}
\end{align}
In the {\em undeformed\/} limit, {\em i.e.} $w=w_0$ and
$\alpha_{\mathcal M}=0$, the theory is the exactly solvable
trigonometric $BC_n$ RSvD \cite{vD} system with a Hamiltonian
$\tilde{\mathcal H}_0$. The corresponding $\phi_0$ becomes the true
ground state wavefunction.
The exact solvability means that $\tilde{\mathcal H}_0$ maps a
Weyl-invariant polynomial in
$\{x_j=\cos\theta_j=\frac{1}{2}(z_j+z_j^{-1})\}$ into another of the
same degree. The $BC_n$ Weyl-invariant polynomials in $\{x_j\}$ are
simply symmetric (under any permutation $j\leftrightarrow k$)
polynomials in $\{x_j\}$ or in $\{z_j+z_j^{-1}\}$.
The eigenfunctions of $\tilde{\mathcal H}_0$ are the $BC$ type Jack
polynomials \cite{jack}.

In the {\em deformed\/} theory, it is straightforward to demonstrate
that $\tilde{\mathcal H}$ maps a Weyl-invariant polynomial in $\{x_j\}$
of degree equal or less than $\mathcal{M}$ into another:
\begin{align}
  \tilde{\mathcal{H}}\,\mathcal{V}_{\mathcal{M}}&\subseteq
  {\mathcal V}_{\mathcal M},
  \qquad\qquad\qquad\qquad\ \ \qquad
  \dim\mathcal{V}_{\mathcal{M}}=\genfrac{(}{)}{0pt}{1}{\mathcal{M}+n}{n},
  \label{subspac2}\\[4pt]
  \mathcal{V}_{\mathcal{M}}&\eqdef\text{Span}\bigl[
  m_{(l_1,\ldots,l_n)}(\{z_j+z_j^{-1}\})\bigm|
  0\leq l_j\leq\mathcal{M},\ 1\leq j\leq n\bigr].
  \label{eq:VtrigBCdef}
\end{align}
This establishes the quasi exact solvability.
The proof that an invariant polynomial is mapped into another goes
almost parallel with the undeformed theory.
One simply has to verify the vanishing of the residues of the simple
poles at $z_j=z_k^{\pm1}, \pm1, \pm q^{1/2},\pm q^{-1/2}$.
The other step is to show that a symmetric polynomial in $\{x_j\}$
with degree $m$ ($\leq\mathcal{M}-1$) is mapped to another with degree
$m+1$, whereas a symmetric polynomial of degree $\mathcal{M}$ remains
with the same degree.
This part goes almost the same as in the single particle case shown
in \cite{deltaqes}, since it is caused by the deformation of $w$,
the single particle interaction part.
The present example is the difference equation version of the QES
theory called trigonometric $BC$ type Inozemtsev system discussed
in section 7 of Sasaki-Takasaki paper \cite{st1}.

%%%%%%%%%%%%%%%%%%%%%%%%%%%%%%%%%%%%%%%%%%%%%%%%%%%%%%%%%%%%%%%
%                                                             %
%  4. Trigonometric $A$ type theory                           %
%                                                             %
%%%%%%%%%%%%%%%%%%%%%%%%%%%%%%%%%%%%%%%%%%%%%%%%%%%%%%%%%%%%%%%
\section{Trigonometric $A$ type theory}
\label{trigA}
\setcounter{equation}{0}

The QES deformation of the trigonometric $A_{n-1}$ RS \cite{RS}
system goes almost parallel with the previous example, or even simpler.
For the $A$-type theory, it is customary to consider $A_{n-1}$ and to
embed all the roots in $\mathbb{R}^{n}$. This is accompanied by the
introduction of one more degree of freedom, $\theta_{n}$ and $p_{n}$.
The genuine $A_{n-1}$ theory corresponds to the relative coordinates
and their momenta, and the extra degree of freedom is the center of
mass coordinate and its momentum.
The Hamiltonian takes the general form \eqref{trigH} with the potential
function $V_j$
\begin{align}
  V_j(z)&\eqdef w(z_j)\prod_{\genfrac{}{}{0pt}{}{k=1}{k\neq j}}^n
  v(z_jz_k^{-1}),\\
  v(y)&\eqdef\frac{1-a_0y}{1-y},\quad w(y)\eqdef 1-a_1y,
  \label{trigAw}\\
  \alpha_{\mathcal{M}}(z)&\eqdef(q^{\mathcal{M}}-1)a_0^{n-1}a_1
  \sum_{j=1}^n\bigl(z_j+z_j^{-1}\bigr).
  \label{trigAcomp}
\end{align}
The undeformed theory, the trigonometric $A_{n-1}$ RS system \cite{RS}
has $w\equiv 1$ and $\alpha_\mathcal{M}=0$.
The main part of the Hamiltonian is factorised as in \eqref{trigfact}
and \eqref{trigAAdag}. The  {\em pseudo ground state\/} wavefunction
$\phi_0$ annihilated by all $\mathcal{A}_j$ reads
\begin{equation}
  \phi_0(z)\eqdef\biggl|\,\prod_{j=1}^n
  \frac{1}{(a_{1}z_j\,;q)_{\infty}}\,\cdot\!\!
  \prod_{1\leq j<k\leq n}
  \frac{(z_jz_k^{-1}\,;q)_{\infty}}{(a_0z_jz_k^{-1}\,;q)_{\infty}}
  \,\biggr|,
\end{equation}
which has no node or singularity in the principal  Weyl alcove of $A_{n-1}$:
\begin{equation}
  PW_T\eqdef\bigl\{\theta\in\mathbb{R}^n\bigm|
  \pi>\theta_1>\theta_2>\cdots >\theta_n>0\bigr\},
  \label{priwaA}
\end{equation}
so long as the parameters are restricted to
\(
  -1<a_0,\, a_1<1
\).
The similarity transformed Hamiltonian $\tilde{\mathcal H}$ in terms of
$\phi_0$ has the same form as \eqref{trigHtilde}, \eqref{trigeigeneq}.

In the {\em undeformed\/} limit, {\em i.e.} $w\equiv 1$ and
$\alpha_{\mathcal M}=0$,
the theory is the exactly solvable trigonometric $A_{n-1}$ RS system
with a Hamiltonian $\tilde{\mathcal{H}}_0$.
The eigenfunctions of $\tilde{\mathcal{H}}_0$ are the well-known Jack
polynomials in $\{z_j\}$ \cite{jack}. Since all the coefficients of the
eigenvalue equation $\tilde{\mathcal{H}}_0\varphi=\mathcal{E}\varphi$
are real, the Jack polynomials in $\{z_j^{-1}\}$ are also eigenfunctions.
In other words $\tilde{\mathcal{H}}_0$ maps a symmetric polynomial in
$\{z_j\}$ into another of the same degree.

The eigenfunctions of the {\em deformed\/} Hamiltonian
$\tilde{\mathcal{H}}$ are still symmetric polynomials in $\{z_j\}$
and  $\{z_j^{-1}\}$, but truncated to the maximal power of $\mathcal{M}$
for each variable, due to the single particle interaction term $w$
\eqref{trigAw} and the compensation term \eqref{trigAcomp}.
This form of the compensation term is necessary for the hermiticity
of the Hamiltonian.
In other words $\tilde{\mathcal H}$ has the invariant subspace
\begin{align}
  \tilde{\mathcal{H}}\,\mathcal{V}_{\mathcal{M}}&\subseteq
  \mathcal{V}_{\mathcal{M}},
  \qquad\qquad\qquad\qquad
  \dim\mathcal{V}_{\mathcal{M}}=\genfrac{(}{)}{0pt}{1}{2\mathcal{M}+n}{n},
  \label{subspac3}\\[4pt]
  \mathcal{V}_{\mathcal{M}}&\eqdef\text{Span}\bigl[
  m_{(l_1,\ldots,l_n)}(\{z_j\})\bigm|
  -\mathcal{M}\leq l_j\leq\mathcal{M},\ 1\leq j\leq n\bigr].
  \label{eq:VtriAdef}
\end{align}
This establishes the quasi exact solvability.
The proof that a symmetric polynomial is mapped into another goes almost
parallel with the undeformed theory. One simply has to verify the
vanishing of the residues of the simple poles at $z_j=z_k$.
The other step is to show that a symmetric polynomial in $\{z_j\}$ with
degree $m$ ($-\mathcal{M}+1\leq m\leq\mathcal{M}-1$) is mapped to
another with degree $m\pm1$, whereas a symmetric polynomial of degree
$\pm\mathcal{M}$ remains with the same degree. This part goes almost
the same as in the single particle case shown in \cite{deltaqes},
since it is caused by the deformation of $w$, the single particle
interaction part.
The present example is the difference equation version of the QES
theory called trigonometric $A$ type Inozemtsev system discussed
in section 8 of Sasaki-Takasaki paper \cite{st1}.

%%%%%%%%%%%%%%%%%%%%%%%%%%%%%%%%%%%%%%%%%%%%%%%%%%%%%%%%%%%%%%%
%                                                             %
%  5. Summary and Comments                                    %
%                                                             %
%%%%%%%%%%%%%%%%%%%%%%%%%%%%%%%%%%%%%%%%%%%%%%%%%%%%%%%%%%%%%%%
\section{Summary and Comments}
\label{summary}
\setcounter{equation}{0}

Quasi exactly solvable multi-particle difference equations are derived
by deforming rational and trigonometric $BC$ type RSvD systems as well
as the trigonometric $A$ type RS systems.
The method of multi-particle deformations is a simple extension of the
single particle case recently developed by one of the authors
\cite{deltaqes}. These examples are the difference equation analogues
of the quasi exactly solvable multi-particle quantum mechanical systems
derived by Sasaki-Takasaki \cite{st1}.

A few comments are in order.
As in the single particle cases, the ranges of the parameters can be
loosened without losing QES.
For example, the six positive parameters in Type II in section 2
\eqref{type2w}, $a_1$, \ldots, $a_6$ could be replaced by three
complex conjugate pairs with positive real parts.
Among the five real parameters $a_1$,\ldots, $a_5$ in
\eqref{trigbcparas}, four could be replaced by two complex conjugate
pairs $b_1$, $b_1^*$, $b_2$, $b_2^*$ with modulus less than unity,
$|b_1|<1$, $|b_2|<1$.

Let us elaborate on the connection with the trigonometric Inozemtsev
systems mentioned at the end of section \ref{trig} and \ref{trigA}.
In fact the examples in section \ref{trig} and \ref{trigA} reduce to the
Inozemtsev systems given in \cite{st1} in a certain limit ($q\to 1$).
Let us introduce a parameter $c$, and set $q=e^{-2/c}$,
$a_0=q^{g}$, $a_1=1-q^{-a/2}$, $\theta=2\theta^{\text{new}}$.
Moreover set
$\mathcal{H}^{\text{new}}
=(a_2a_3a_4a_5q^{-1})^{-1/2}a_0^{-(n-1)}\mathcal{H}$ and
$(a_2,a_3,a_4,a_5)=(q^{g_1},q^{g_2+1/2},-q^{g'_1},-q^{g'_2+1/2})$
for the $BC$ type theory,
$\mathcal{H}^{\text{new}}=a_0^{-(n-1)/2}\mathcal{H}$
for the $A$ type theory (see \cite{os6}).
Then in the $c\to\infty$ limit, the Hamiltonian
$\frac12c^2\mathcal{H}^{\text{new}}$ reduces to the trigonometric
Inozemtsev systems discussed in section 7 and 8 of \cite{st1}.

The corresponding statement for the rational theory is somehow complicated
and it requires a double limit.
For the parameters in section \ref{rat}, let us set 
$(a_3,a_4,a_5,a_6)=(
\frac{c}{\omega}\sqrt{\frac{\omega_1}{a}}
+\frac{1}{2a}\frac{\omega_1}{\omega}(\frac{\omega_1}{\omega}-1),
-\frac{c}{\omega}\sqrt{\frac{\omega_1}{a}}
+\frac{1}{2a}\frac{\omega_1}{\omega}(\frac{\omega_1}{\omega}-1),
g_1,g_2+\frac12)$ and $x_j=cx^{\text{new}}_j$.
Moreover set 
$\tilde{\mathcal{H}}^{\text{new}}=4(a_2a_3a_4)^{-1}\tilde{\mathcal{H}}$
and $a_2=c^2/\omega_1$ for the type I theory,
$\tilde{\mathcal{H}}^{\text{new}}=4(a_1a_2a_3a_4)^{-1}\tilde{\mathcal{H}}$
and $a_1=a_2=2c^2/\omega_1$ for the type II theory (see \cite{os6}).
Then a double limit
$\lim\limits_{\omega_1\to\infty}\bigl(\lim\limits_{c\to\infty}
\frac12c^2\tilde{\mathcal{H}}^{\text{new}}\bigr)$
gives the Hamiltonian of the rational Inozemtsev system discussed in section 6
of \cite{st1}.

It is interesting to note that the weight function $\phi_0^2(x)$ for 
the polynomial eigenfunctions $\{P_{\mathcal M}(x)\}$ is the zero mode
(stationary distribution) of the corresponding deformed Fokker-Planck
equation \cite{hs}.

%%%%%%%%%%%%%%%%%%%%%%%%%%%%%%%%%%%%%%%%%%%%%%%%%%%%%%%%%%%%%%%
%                                                             %
%  Acknowledgments                                            %
%                                                             %
%%%%%%%%%%%%%%%%%%%%%%%%%%%%%%%%%%%%%%%%%%%%%%%%%%%%%%%%%%%%%%%
\section*{Acknowledgements}

This work is supported in part by Grants-in-Aid for Scientific
Research from the Ministry of Education, Culture, Sports, Science and
Technology, No.18340061 and No.19540179.

%%%%%%%%%%%%%%%%%%%%%%%%%%%%%%%%%%%%%%%%%%%%%%%%%%%%%%%%%%%%%%%
%                                                             %
%  References                                                 %
%                                                             %
%%%%%%%%%%%%%%%%%%%%%%%%%%%%%%%%%%%%%%%%%%%%%%%%%%%%%%%%%%%%%%%

\end{document}